\documentclass{article}
\usepackage{spconf,amsmath,graphicx, mathtools}
\usepackage{amssymb}
\usepackage{booktabs}
\usepackage{subfig}
\usepackage[english]{babel}
\usepackage{xcolor}
\usepackage{makecell}
\usepackage{varwidth}
\usepackage{xcolor}
\usepackage{paralist}
\usepackage{bbm}
\usepackage{multirow}
\usepackage{hyperref}
\usepackage{breqn}
\usepackage{longtable}
\usepackage{mathtools}
\usepackage{algorithm}
\usepackage{algpseudocode}
\usepackage{paralist}
\usepackage{microtype}
\usepackage[detect-all]{siunitx}
\usepackage{cleveref}
\usepackage{lscape} 
\usepackage{upgreek}
\usepackage{multirow}
\usepackage{booktabs}
\usepackage{graphicx}

\title{On Out-of-Distribution Detection for Audio with Deep Nearest Neighbors}
\name{Zaharah Bukhsh, Aaqib Saeed\thanks{The icons used in the figure are from TheNounProject.}}
\address{
Eindhoven University of Technology, Eindhoven, The Netherlands\\
}

\begin{document}
\maketitle

\begin{abstract}
Out-of-distribution (OOD) detection is concerned with identifying data points that do not belong to the same distribution as the model's training data. For the safe deployment of predictive models in a real-world environment, it is critical to avoid making confident predictions on OOD inputs as it can lead to potentially dangerous consequences. However, OOD detection largely remains an under-explored area in the audio (and speech) domain. This is despite the fact that audio is a central modality for many tasks, such as speaker diarization, automatic speech recognition, and sound event detection. To address this, we propose to leverage feature-space of the model with deep \textit{k}-nearest neighbors to detect OOD samples. We show that this simple and flexible method effectively detects OOD inputs across a broad category of audio (and speech) datasets. Specifically, it improves the false positive rate (FPR@TPR95) by $17\%$ and the AUROC score by $7\%$ than other prior techniques.
\end{abstract}

\keywords{out-of-distribution, audio, speech, uncertainty estimation, deep learning, nearest neighbors}

\section{Introduction}
\label{sec:introduction}
Out-of-distribution (OOD) detection is the task of identifying inputs that are not drawn from the same distribution as the training data or are not truly representative of them. Neural networks are known to produce overconfident scores even for samples that do not belong to the training distribution~\cite{liu2020energy}. This is a challenging problem for deploying machine learning in safety-critical applications, where making confident predictions on OOD inputs can lead to potentially dangerous consequences. Besides the capability to generalize well for samples from the familiar distribution, a robust machine learning model should be aware of uncertainty stemming from unknown examples. It is an important competency for real-world applications, where the distribution of data can change over time or vary across different user groups.

A broad range of approaches has been proposed to tackle the OOD detection issue and develop reliable methods that successfully detect in-distribution (ID) and OOD inputs. A set of common techniques is to deduce uncertainty measurements around predictions of the neural network based on model outputs~\cite{liu2020energy,hendrycks2016baseline,liang2017enhancing,lakshminarayanan2017simple}, feature space~\cite{sun2022out, lee2018simple}, and gradient norms~\cite{huang2021importance}. Similarly, distance-based methods~\cite{sun2022out} has also gained significant attention recently for identifying OOD inputs with promising capabilities. Distance-based methods leverage representations extracted from a pre-trained model and act on the assumption that out-of-distribution test samples are isolated from the ID data. Nevertheless, OOD detection is severely understudied in the audio domain, although audio recognition models are being widely deployed in real-world settings. As well as, audio is an important modality for many tasks, such as speaker diarization, automatic speech recognition, and sound event detection. The prior works mainly focus on vision tasks raising an important question about the efficacy and applicability of existing methods to audio and speech. 

Our work follows the same intuition as of the distance-based method~\cite{sun2022out}, and we aim to explore the richness of model representation space to derive a meaningful signal that can help solve the task of OOD detection. Formally, we propose a simple yet effective system for \textit{out-of-distribution detection for audio} inputs with deep \textit{k}-nearest neighbors. In particular, we leverage nearest neighbor distance centered on a non-parametric approach without making strong distributional assumptions regarding underlying embedding space. To identify OOD samples, we extract embedding for a test input, compute its distance to \textit{k}-nearest neighbors in the training set and use a threshold to flag the input, i.e., a sample far away in representation space is more likely to be OOD.

We demonstrate the effectiveness of \textit{k}NN-based approach on a broad range of audio recognition tasks and different neural network architectures and provide an extensive comparison with both recent and classical approaches as baselines. Importantly, we note that to the best of our knowledge, we make a first attempt at studying out-of-distribution detection and setting up a benchmark for audio across a variety of datasets ranging from keyword spotting and emotion recognition to environmental sounds and more. Empirically, we show that for a MobileNet~\cite{howard2017mobilenets} model (trained on in-distribution data of human vocal sounds), the non-parametric nearest neighbor method improves FPR@TPR95 by $17$\% and AUROC scores by $7$\% than approaches that leverage output or gradient space of the model. 

\begin{figure*}[t]
    \centering
    \includegraphics[width=0.86\textwidth]{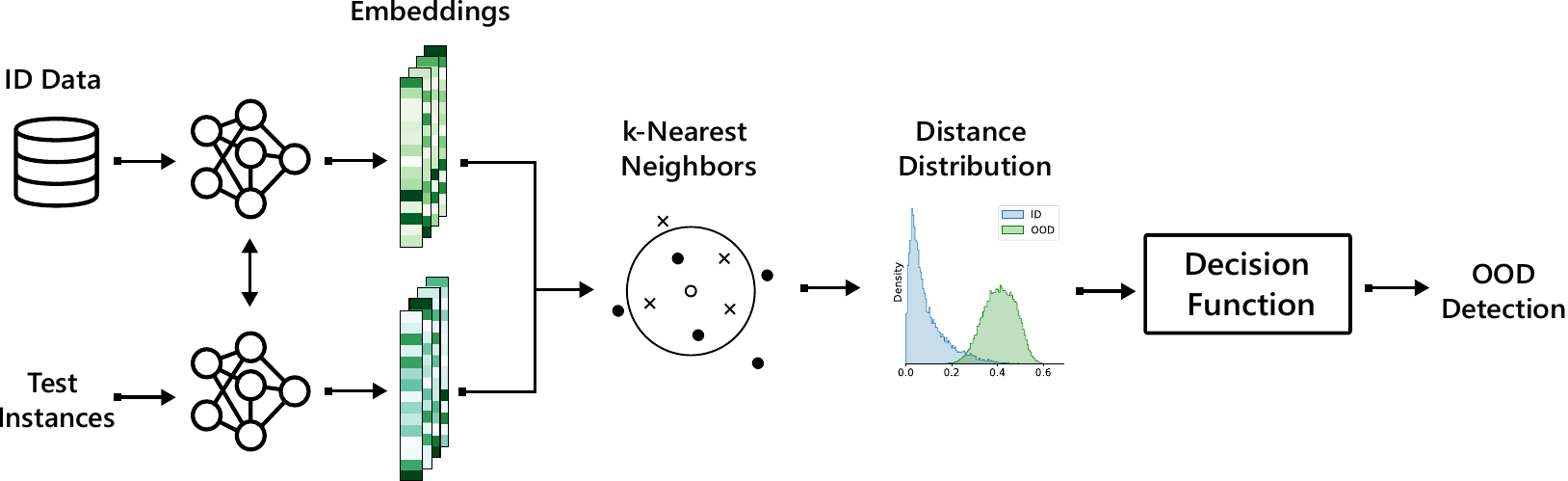}
    \caption{Overview of the \textit{k}-deep nearest neighbors approach for leveraging embedding space to detect out-of-distribution samples.}
    \label{fig:audio_ood_overivew}
\end{figure*}

\section{Approach}
\subsection{Preliminaries}
\noindent \textbf{Learning Regime} We focus on supervised learning regime, specifically, multi-class classification tasks, where, $\mathcal{X}$ and $\mathcal{Y} = \{ 1, \ldots C\}$ denote input and label spaces, respectively. A classifier $f_{\theta}(\cdot)$ utilize training set $\mathcal{D}_{p} = \{ (x_i, y_i) \}_{i=1}^{M}$, which is drawn i.i.d from the joint distribution $\mathcal{P}$ defined over $\mathcal{X} \times \mathcal{Y}$. The deep model $f_{\theta}(x): \mathcal{X} \longrightarrow \mathbb{R}^{|\mathcal{Y}|}$ minimizes negative log-likelihood (or similar) objective with back-propagation to produce logits that are then translated to predicted labels being assigned to the input samples. \\

\noindent \textbf{Problem Formulation} Out-of-distribution (OOD) detection is generally formulated as a binary classification problem with the objective of identifying samples from unknown data distribution at inference time. For instance, samples from an unrelated distribution whose label set does not overlap with the task labels (of a trained deep model) should be deferred instead of producing incorrect prediction~\cite{liu2020energy}. Formally, given a pre-trained classifier $f_{\theta}(\cdot)$ that is learned to solve a task $t$ using data $\mathcal{D}_{p}$ from in-domain data distribution, the aim of OOD is to have a decision function:
\begin{align*}
	\mathcal{U}_{\gamma}(\*x)=\begin{cases} 
      \text{ID} & \mathcal{H}(x)\ge \gamma \\
      \text{OOD} & \mathcal{H}(x) < \gamma 
  \end{cases}
\end{align*}
\noindent that flags whether a sample $x \in \mathcal{X}$ does not belongs to $\mathcal{D}_{p}$. $\gamma$ represents a threshold that is chosen such that a large fraction (e.g., $95\%$) of ID samples are correctly identified. The domain of OOD detection is concerned with the development of a scoring function $\mathcal{H}$ that captures the uncertainty of data being outside of the training data distribution. Previous approaches largely relies on output~\cite{liu2020energy,hendrycks2016baseline,lakshminarayanan2017simple,liang2017enhancing}, feature~\cite{lee2018simple} and gradient~\cite{huang2021importance} spaces of the model, with~\cite{sun2022out} proposing to leverage nearest neighbors in the feature space to determine uncertainty. Along similar lines, we propose to leverage the non-parametric nearest neighbors approach to detect OOD samples in audio, as we describe in the subsequent section. 

\begin{table*}[t]
\small
\caption{\small{OOD detection results in comparison with other strong approaches on different audio datasets. All results are based on a MobileNet~\cite{howard2017mobilenets} (YAMNet) model trained only on ID data (i.e., MSWC (Micro-EN) and Vocalsound). $\uparrow$ and $\downarrow$ indicate larger and smaller values are better, respectively. All values are percentages.}}
\centering
\subfloat[Vocalsound]{
\label{tab:vocalsound}
\begin{tabular}{@{}lcccccccc@{}}
    \toprule
    \multirow{3}{*}{\textbf{OOD Dataset}} & \multicolumn{8}{c}{\textbf{Method}}                           \\
     & \multicolumn{2}{c}{\textbf{MSP}} & \multicolumn{2}{c}{\textbf{ODIN}} & \multicolumn{2}{c}{\textbf{GradNorm}} & \multicolumn{2}{c}{\textbf{\textit{k}NN}} \\
     & FPR95$\downarrow$ & AUROC$\uparrow$ & FPR95$\downarrow$ & AUROC$\uparrow$ & FPR95$\downarrow$ & AUROC$\uparrow$ & FPR95$\downarrow$ & AUROC$\uparrow$ \\ \midrule
    MSWC (Micro-EN)                       &   82.44    &    76.41   &     87.79  &   75.80    &    83.73   &   76.81    &    53.70   &   89.61    \\
    Voxforge                              &   39.84    &    94.22   &      31.45 &   95.33   &   31.48    &   95.10    &   39.43    &   94.28    \\
    CREMA-D                               &   51.74    &   91.64    &     46.92  &   93.03    &   46.34    &   92.48   &    28.28   &  95.53     \\
    ESC-50                                &    58.25   &   85.51    &     53.75  &   86.75    &    54.75   &   86.17    &    49.0   &   91.33    \\
    MSWC (Micro-ES)                       &   81.58    &    76.86   &     88.23  &    75.21   &    83.64   &   76.84    &   46.20    &   90.91    \\
    SpeechCommands                            &    79.59   &   78.55      &     83.72    &    76.67   &   80.51    &    78.55   &   48.96    &    91.49   \\
    FluentSpeech                          &   57.58    &   90.49    &   55.87    &   91.52      &   52.41   &   91.38    &   42.05    &  93.65     \\ \hline
    Average                          &     64.43  &   84.85    &   63.96    &    84.90   &   61.84    &    85.33  &   \textbf{43.94}    &     \textbf{92.40}  \\\bottomrule
\end{tabular}%
}

\subfloat[MSWC (Micro-EN)]{
\label{tab:mswc_en}
\begin{tabular}{@{}lcccccccc@{}}
    \toprule
    \multirow{3}{*}{\textbf{OOD Dataset}} & \multicolumn{8}{c}{\textbf{Method}}                           \\
     & \multicolumn{2}{c}{\textbf{MSP}} & \multicolumn{2}{c}{\textbf{ODIN}} & \multicolumn{2}{c}{\textbf{GradNorm}} & \multicolumn{2}{c}{\textbf{\textit{k}NN}} \\
     & FPR95$\downarrow$ & AUROC$\uparrow$ & FPR95$\downarrow$ & AUROC$\uparrow$ & FPR95$\downarrow$ & AUROC$\uparrow$ & FPR95$\downarrow$ & AUROC$\uparrow$ \\ \midrule
    Vocalsound                        &   36.23    &    95.53   &    2.87   &    99.26   &    12.20    &  97.67    &   4.65    &   98.80    \\
    Voxforge                              &  31.24     &  95.92    &  2.40     &    99.40   &   11.26     &    97.81    &   2.73   &   98.94   \\
    CREMA-D                               &   31.17    &    95.79   &  2.51     &  99.39     &  11.57     &   97.71    &    2.63   &    99.25   \\
    ESC-50                                &   34.0    &      95.91 &      5.50 &     98.89  &    9.75   &    98.0   &   9.75    &    98.59   \\
    MSWC (Micro-ES)                       &   67.33    &   87.81    &     56.21  &   89.42    &    56.39   &   88.74    &    39.37   &  93.11     \\
    SpeechCommands                            &   73.01    &   76.12      &   63.44    &  77.58    &   64.83    &  76.81     &    52.29   &     84.14  \\
    FluentSpeech                          &    42.82   &    94.40   &      7.30  &   98.47    &    21.91   &   96.27    &    7.62   &  98.33      \\ \hline
    Average                          &   45.11    &    91.64   &    20.03   &   94.62    &    26.84   &   93.29    &    \textbf{17.0}   &   \textbf{95.88}    \\\bottomrule
\end{tabular}%
}
\label{tab:main_table}
\end{table*}

\subsection{OOD Detection with Deep \textit{k}-Nearest Neighbor}
We aim to exploit the representation space of a pre-trained neural network for detecting out-of-distribution samples with \textit{k}-nearest neighbor approach. We provide a high-level illustration of our approach in Figure~\ref{fig:audio_ood_overivew}. The key driving factor behind distance-based non-parametric methods is that distances in the embedding space provide a meaningful way to compare data from distributions. Hence, they can be utilized to identify OOD samples as ID samples are closer to each other in the feature space as compared to OOD data points. Inspired by the simplicity and success of the deep nearest neighbor method for OOD detection in vision domain~\cite{sun2022out}, we propose to leverage and study whether we can use it to reliably detect samples different than the ID training set in audio (and speech) domain. 

Given a pre-trained classification model $f_{\theta}(x): \mathcal{X} \longrightarrow \mathbb{R}^{|\mathcal{Y}|}$, we extract normalized representations (features or embeddings) $z = \frac{\psi(x)}{|| \psi(x) ||_{2}}$ from penultimate layer of the model, where $\psi$ can be seen as a feature extractor. With $\mathcal{Z}_{m} = (z_1, \ldots, z_n)$ be the embedding vectors from an ID training set and $z^{*}$ be an embedding for a test sample. We compute Euclidean distance of test input $||z_{i} - z^{*} ||_{2}$ with each example in $\mathcal{Z}_{m}$ and reorder element in $\mathcal{Z}_{m}$ based on increasing distance. Finally, we use a decision function from~\cite{sun2022out} to check if sample is OOD as: $\mathcal{H}(z^{*}, k) = \mathbf{1}\{-d_k(z^{*}) \geq \gamma\}$, where, $d_{k} = ||z_{k} - z^{*}||_{2}$ denotes distance to $\textit{k}$-th nearest neighbor and $\mathbf{1}\{\cdot\}$ is an indicator function. In practice, the threshold $\gamma$ can be chosen to correctly classify a large percentage (e.g., $95\%$) of ID samples. It is important to note that picking the optimal value of $\gamma$ does not depend on OOD data. 

There are several advantages of using deep nearest neighbor over other methods for out-of-distribution detection. First, it is scalable and can be used with large data sets using an efficient similarity search library, such as Faiss~\cite{johnson2019billion}. Second, it is easy to use as it does not require access to out-of-distribution data for defining threshold. Third, it is model-agnostic in the sense that we can use it with a variety of neural network architectures and different training regimes (i.e., both supervised and unsupervised). Finally, it also offers an interpretability into the process of identifying OOD samples by letting human operator listen to the $k$ closest training samples to the test one. 
\begin{figure}[!htbp]
    \centering
    \subfloat{
        \includegraphics[width=0.38\columnwidth]{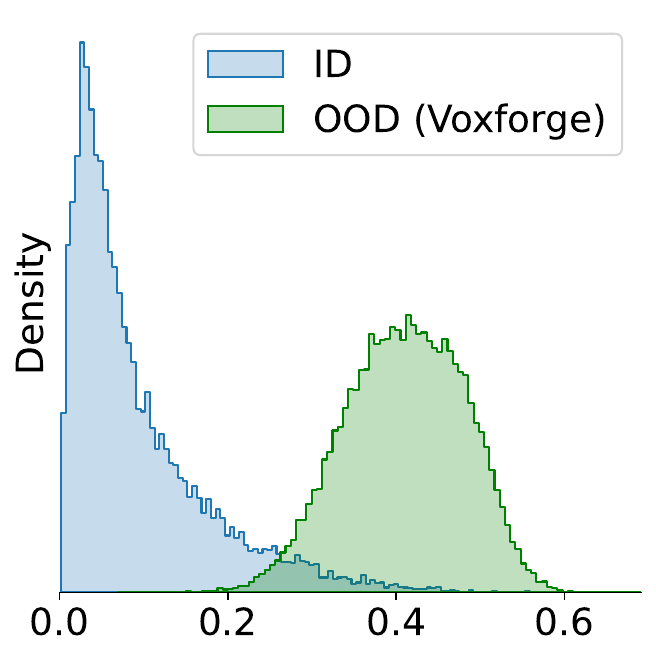}
    }
    \subfloat{
        \includegraphics[width=0.38\columnwidth]{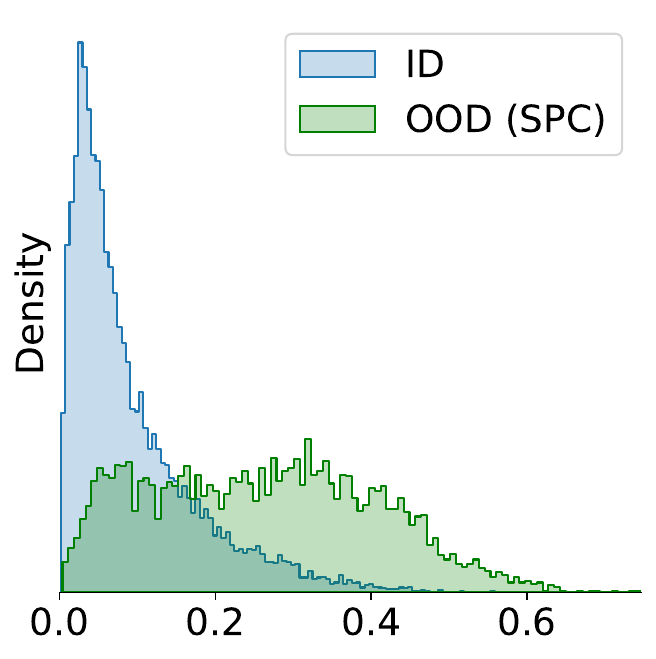}
    }  \\
    \subfloat{
        \includegraphics[width=0.38\columnwidth]{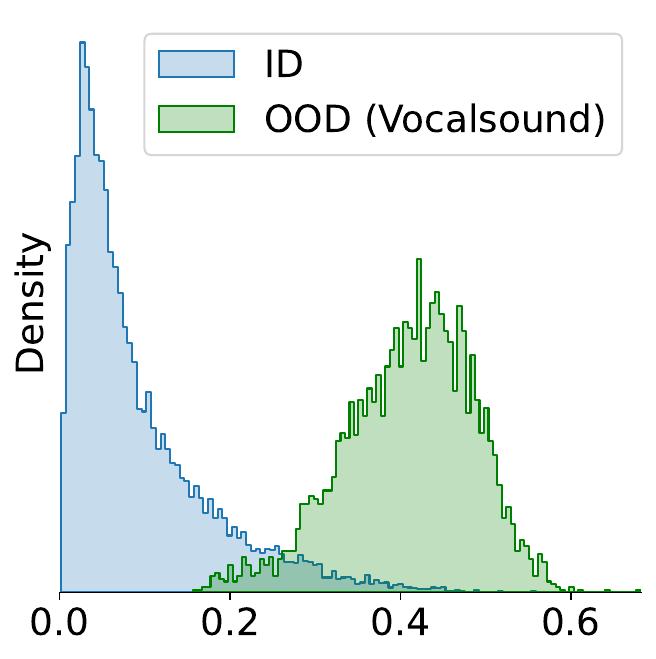}
    }  
    \subfloat{
        \includegraphics[width=0.38\columnwidth]{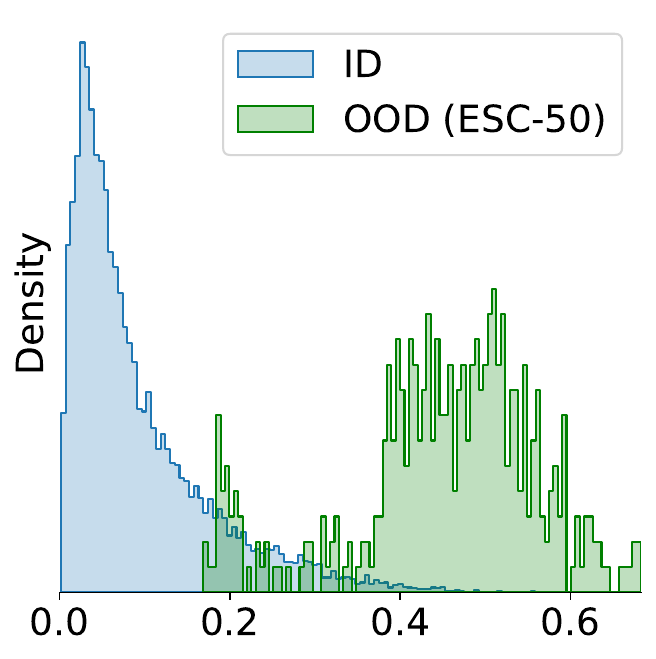}
    }  
    \caption{\small{Distribution of the \textit{k}NN distance with the normalized features from MobileNet. The ID data is MSWC (Micro-EN).}}
    \label{fig:distance_visual}
\end{figure}

\section{Experiments}  
\label{sec:experiments}

\subsection{Datasets and Evaluation}
We use MSCW (Micro-EN)~\cite{mazumder2021multilingual} and Vocalsound~\cite{gong_vocalsound} as in-distribution datasets, and they also act as OOD for each other. The former is curated for learning keyword spotting models with $31$ classes, and the latter is related to human vocal sound monitoring (e.g., coughing and sneezing) with $6$ classes. We use standard training, validation, and test splits as provided with these datasets with audio sampled at $16$kHz. For the OOD test dataset, we employ six additional datasets to cover wide range of audio and speech, including Voxforge~\cite{maclean2018voxforge} (spoken language identification, $6$ classes), CREMA-D~\cite{cao2014crema} (emotion recognition, $6$ classes), ESC-50~\cite{piczak2015dataset} (environmental sound classification, $50$ classes), MSWC (Micro-ES)~\cite{mazumder2021multilingual} (Spanish keyword spotting, $20$ classes), SpeechCommands~\cite{warden2018speech} (spoken commands detection, $12$ classes), and FluentSpeech~\cite{lugosch2019speech} (action recognition, $6$ classes). We note that the model is \underline{not} exposed to any OOD data during its training phase, and test sets of OOD datasets are used to report results. For evaluation, we closely follow prior work on OOD detection in vision~\cite{sun2022out, liu2020energy, huang2021importance} and compute the false positive rate (FPR95) of OOD examples when the true positive rate of in-distribution examples is set at $95$\%. Further, we also report the area under the receiver operating characteristic curve (AUROC). 

\subsection{Models and Implementation Details}
We use MobileNet~\cite{howard2017mobilenets} (YAMNet) architecture for training audio classification models from scratch with negative log-likelihood loss. The model input is  log-compressed Mel-filterbanks with a window size of $25$ms, a hop size of $10$ms, and $N=64$ Mel-spaced frequency bins in the range $60$–$7800$Hz for $T=98$ frames, corresponding to $980$ms. We use Adam~\cite{kingma2014adam} optimizer with a learning rate of $0.001$ for training the model for $100$ epochs and batch size of $128$. The penultimate layer's feature dimension is $1024$, where the nearest neighbor search is performed. We also explore the efficacy of our approach with other architecture for which we use EfficientNet-B0~\cite{tan2019efficientnet}, where the training configuration is the same as mentioned earlier, and the dimensionality of the penultimate features is $1280$. For efficient nearest neighbors search, we use Faiss~\cite{johnson2019billion} with Euclidean distance and set $k=5$ and $k=100$ for in-domain MSWC (Micro-EN) and Vocalsound datasets, respectively. Likewise, for other traditional non-parametric methods( see Table~\ref{tab:trd_methods_table}) we use  scikit-learn~\cite{scikit-learn} and PyOD~\cite{zhao2019pyod}. Specifically, we use $100$ base estimators for IForest~\cite{liu2008isolation}, apply RBF kernel in OCSVM with $nu=0.25$~\cite{scholkopf2001estimating}, $10$ number of bins for LODA~\cite{pevny2016loda}, $k=20$ for LOF~\cite{breunig2000lof}, and use $128$ components in PCA~\cite{shyu2003novel}. 

\subsection{Results and Analysis}
\begin{table}[!t]
\small
\caption{\small{OOD detection results on EfficientNet-B0.}}
\label{tab:avg_efnet_table}
\centering
\begin{tabular}{@{}lllll@{}}
\toprule
\multicolumn{1}{c}{\multirow{2}{*}{\textbf{Method}}} & \multicolumn{2}{c}{\textbf{Vocalsound}}               & \multicolumn{2}{c}{\textbf{MSWC (Micro-EN)}}          \\
\multicolumn{1}{c}{}                                 & \multicolumn{1}{c}{FPR95$\downarrow$} & \multicolumn{1}{c}{AUROC$\uparrow$} & \multicolumn{1}{c}{FPR95$\downarrow$} & \multicolumn{1}{c}{AUROC$\uparrow$} \\ \midrule
MSP     & 35.93 & 94.06 & 45.61 &  92.19 \\
ODIN     & 16.77  & 96.53 & 41.01 & 92.17 \\
GradNorm & 24.15 & 94.86 & 32.39 & 93.27 \\ \hline
\textit{k}NN     &  \textbf{16.71} &  \textbf{96.87} & \textbf{19.63} & \textbf{95.20} \\ \bottomrule
\end{tabular}
\end{table}

In this section, we investigate the feasibility of \textit{k}-deep nearest neighbors method for detecting out-of-distribution samples. We use MobileNet~\cite{howard2017mobilenets} models trained solely on in-distribution data to highlight the feasibility of our method, unless stated otherwise. In Table~\ref{tab:main_table}, we compare results with competitive methods leveraging output and even gradient space of the model. When using Vocalsound as ID data (test set accuracy of $90.75$\%), where the task involves human vocal sound recognition, the \textit{k}NN improves FPR95 by $17$\% while being extremely efficient as compared to more sophisticated methods as ODIN and GradNorm, which are computationally intensive as they require computing per-sample gradients. Further, as \textit{k}NN has an entire training set embeddings at its disposal, it works effectively than traditional methods as it compares distance of a test input to rest of the samples to determine the OOD status.

Likewise, when a keyword spotting model trained on MSWC (Micro-EN) (with test set accuracy of $90.05$\%) is used we notice ODIN performs relatively better but \textit{k}NN provides better FPR with an improvement of $3\%$. We further provide distance distribution in Figure~\ref{fig:distance_visual} of demonstrating clear separation of ID and OOD distributions. Based on these results, we notice that the nature of ID data has a strong impact on the performance of OOD detection. For instance, MSWC contains audio of spoken words in general of one second in length as compared to variable length audio in Vocalsound dataset. Hence, the high FPR we observe on SpeechCommands (also see distance overlap in Figure~\ref{fig:distance_visual}) and MSWC (Micro-ES), i.e., when ID and OOD datasets have similar task characteristics, OOD identification is more challenging.  

We also evaluate \textit{k}NN on an alternative neural network architecture, EfficientNet-B0~\cite{tan2019efficientnet}. In Table~\ref{tab:avg_efnet_table}, we report averaged evaluation metrics across datasets mentioned in Table~\ref{tab:main_table} and follow same experimental setting. We see that non-parametric \textit{k}NN is largely effective and outperforms existing techniques. Further, as we can see in Table~\ref{tab:trd_methods_table}, the proposed \textit{k}NN-based method outperforms all the other classical techniques in terms of both AUROC and FPR95 by a significant margin, while using same embeddings used for training of these other classical models.

\begin{table}[!t]
\small
\caption{\small{\textit{k}NN comparison with classical non-parametric methods. The MobileNet model is trained on Vocalsound. The results are averaged across all test OOD datasets as mentioned in Table~\ref{tab:vocalsound}.}}
\label{tab:trd_methods_table}
\centering
\begin{tabular}{@{}lcc@{}}
\toprule & FPR95 $\downarrow$ & AUROC $\uparrow$ \\ \midrule
IForest &  97.21     &   52.09   \\
OCSVM   &    61.24   &    76.70   \\
LODA    &   97.75    &    32.69   \\
LOF     &   70.74    &   85.15    \\
PCA     & 93.92      &  51.92 \\ \hline
\textit{k}NN     &    \textbf{43.94}   &    \textbf{92.40}   \\ \bottomrule
\end{tabular}
\end{table}

\section{Conclusions} 
\label{sec:conclusions}
We present a robust technique for out-of-distribution detection in the audio domain with deep \textit{k}-nearest neighbors. Our approach leverages a non-parametric nearest-neighbor distance without making strong distributional assumptions regarding the underlying embedding space of a pre-trained neural network model. We demonstrate the effectiveness of \textit{k}NN in OOD detection for audio (and speech) across a wide variety of tasks and model architectures. Our approach has the potential to improve the safety of audio models deployed in real-world settings by reducing the risk of making overconfident predictions on inputs that are not related to the task at hand. 

\bibliographystyle{IEEEbib}
\bibliography{main}

\begin{thebibliography}{10}

\bibitem{liu2020energy}
Weitang Liu, Xiaoyun Wang, John Owens, and Yixuan Li,
\newblock ``Energy-based out-of-distribution detection,''
\newblock {\em Advances in Neural Information Processing Systems}, vol. 33, pp.
  21464--21475, 2020.

\bibitem{hendrycks2016baseline}
Dan Hendrycks and Kevin Gimpel,
\newblock ``A baseline for detecting misclassified and out-of-distribution
  examples in neural networks,''
\newblock {\em arXiv preprint arXiv:1610.02136}, 2016.

\bibitem{liang2017enhancing}
Shiyu Liang, Yixuan Li, and Rayadurgam Srikant,
\newblock ``Enhancing the reliability of out-of-distribution image detection in
  neural networks,''
\newblock {\em arXiv preprint arXiv:1706.02690}, 2017.

\bibitem{lakshminarayanan2017simple}
Balaji Lakshminarayanan, Alexander Pritzel, and Charles Blundell,
\newblock ``Simple and scalable predictive uncertainty estimation using deep
  ensembles,''
\newblock {\em Advances in neural information processing systems}, vol. 30,
  2017.

\bibitem{sun2022out}
Yiyou Sun, Yifei Ming, Xiaojin Zhu, and Yixuan Li,
\newblock ``Out-of-distribution detection with deep nearest neighbors,''
\newblock {\em arXiv preprint arXiv:2204.06507}, 2022.

\bibitem{lee2018simple}
Kimin Lee, Kibok Lee, Honglak Lee, and Jinwoo Shin,
\newblock ``A simple unified framework for detecting out-of-distribution
  samples and adversarial attacks,''
\newblock {\em Advances in neural information processing systems}, vol. 31,
  2018.

\bibitem{huang2021importance}
Rui Huang, Andrew Geng, and Yixuan Li,
\newblock ``On the importance of gradients for detecting distributional shifts
  in the wild,''
\newblock {\em Advances in Neural Information Processing Systems}, vol. 34, pp.
  677--689, 2021.

\bibitem{howard2017mobilenets}
Andrew~G Howard, Menglong Zhu, Bo~Chen, Dmitry Kalenichenko, Weijun Wang,
  Tobias Weyand, Marco Andreetto, and Hartwig Adam,
\newblock ``Mobilenets: Efficient convolutional neural networks for mobile
  vision applications,''
\newblock {\em arXiv preprint arXiv:1704.04861}, 2017.

\bibitem{johnson2019billion}
Jeff Johnson, Matthijs Douze, and Herv{\'e} J{\'e}gou,
\newblock ``Billion-scale similarity search with gpus,''
\newblock {\em IEEE Transactions on Big Data}, vol. 7, no. 3, pp. 535--547,
  2019.

\bibitem{mazumder2021multilingual}
Mark Mazumder, Sharad Chitlangia, Colby Banbury, Yiping Kang, Juan~Manuel Ciro,
  Keith Achorn, Daniel Galvez, Mark Sabini, Peter Mattson, David Kanter,
  et~al.,
\newblock ``Multilingual spoken words corpus,''
\newblock in {\em Thirty-fifth Conference on Neural Information Processing
  Systems Datasets and Benchmarks Track (Round 2)}, 2021.

\bibitem{gong_vocalsound}
Yuan Gong, Jin Yu, and James Glass,
\newblock ``Vocalsound: A dataset for improving human vocal sounds
  recognition,''
\newblock in {\em ICASSP 2022 - 2022 IEEE International Conference on
  Acoustics, Speech and Signal Processing (ICASSP)}, 2022, pp. 151--155.

\bibitem{maclean2018voxforge}
Ken MacLean,
\newblock ``Voxforge,''
\newblock {\em Ken MacLean.[Online]. Available: http://www. voxforge.
  org/home.[Acedido em 2012]}, 2018.

\bibitem{cao2014crema}
Houwei Cao, David~G Cooper, Michael~K Keutmann, Ruben~C Gur, Ani Nenkova, and
  Ragini Verma,
\newblock ``{CREMA-D}: Crowd-sourced emotional multimodal actors dataset,''
\newblock {\em IEEE transactions on affective computing}, vol. 5, no. 4, pp.
  377--390, 2014.

\bibitem{piczak2015dataset}
Karol~J. Piczak,
\newblock ``{ESC}: {Dataset} for {Environmental Sound Classification},''
\newblock in {\em Proceedings of the 23rd {Annual ACM Conference} on
  {Multimedia}}. pp. 1015--1018, {ACM Press}.

\bibitem{warden2018speech}
Pete Warden,
\newblock ``Speech commands: A dataset for limited-vocabulary speech
  recognition,''
\newblock {\em arXiv preprint arXiv:1804.03209}, 2018.

\bibitem{lugosch2019speech}
Loren Lugosch, Mirco Ravanelli, Patrick Ignoto, Vikrant~Singh Tomar, and Yoshua
  Bengio,
\newblock ``Speech model pre-training for end-to-end spoken language
  understanding,''
\newblock {\em arXiv preprint arXiv:1904.03670}, 2019.

\bibitem{kingma2014adam}
Diederik~P Kingma and Jimmy Ba,
\newblock ``Adam: A method for stochastic optimization,''
\newblock {\em arXiv preprint arXiv:1412.6980}, 2014.

\bibitem{tan2019efficientnet}
Mingxing Tan and Quoc Le,
\newblock ``Efficientnet: Rethinking model scaling for convolutional neural
  networks,''
\newblock in {\em International conference on machine learning}. PMLR, 2019,
  pp. 6105--6114.

\bibitem{scikit-learn}
F.~Pedregosa, G.~Varoquaux, A.~Gramfort, V.~Michel, B.~Thirion, O.~Grisel,
  M.~Blondel, P.~Prettenhofer, R.~Weiss, V.~Dubourg, J.~Vanderplas, A.~Passos,
  D.~Cournapeau, M.~Brucher, M.~Perrot, and E.~Duchesnay,
\newblock ``Scikit-learn: Machine learning in {P}ython,''
\newblock {\em Journal of Machine Learning Research}, vol. 12, pp. 2825--2830,
  2011.

\bibitem{zhao2019pyod}
Yue Zhao, Zain Nasrullah, and Zheng Li,
\newblock ``Pyod: A python toolbox for scalable outlier detection,''
\newblock {\em Journal of Machine Learning Research}, vol. 20, no. 96, pp.
  1--7, 2019.

\bibitem{liu2008isolation}
Fei~Tony Liu, Kai~Ming Ting, and Zhi-Hua Zhou,
\newblock ``Isolation forest,''
\newblock in {\em 2008 eighth ieee international conference on data mining}.
  IEEE, 2008, pp. 413--422.

\bibitem{scholkopf2001estimating}
Bernhard Sch{\"o}lkopf, John~C Platt, John Shawe-Taylor, Alex~J Smola, and
  Robert~C Williamson,
\newblock ``Estimating the support of a high-dimensional distribution,''
\newblock {\em Neural computation}, vol. 13, no. 7, pp. 1443--1471, 2001.

\bibitem{pevny2016loda}
Tom{\'a}{\v{s}} Pevn{\`y},
\newblock ``Loda: Lightweight on-line detector of anomalies,''
\newblock {\em Machine Learning}, vol. 102, no. 2, pp. 275--304, 2016.

\bibitem{breunig2000lof}
Markus~M Breunig, Hans-Peter Kriegel, Raymond~T Ng, and J{\"o}rg Sander,
\newblock ``Lof: identifying density-based local outliers,''
\newblock in {\em Proceedings of the 2000 ACM SIGMOD international conference
  on Management of data}, 2000, pp. 93--104.

\bibitem{shyu2003novel}
Mei-Ling Shyu, Shu-Ching Chen, Kanoksri Sarinnapakorn, and LiWu Chang,
\newblock ``A novel anomaly detection scheme based on principal component
  classifier,''
\newblock Tech. {R}ep., Miami Univ Coral Gables Fl Dept of Electrical and
  Computer Engineering, 2003.

\end{thebibliography}
\end{document}